\documentstyle[aps,prd]{revtex}
\input psfig.tex
\def\slash#1{\setbox0=\hbox{$#1$}#1\hskip-\wd0\hbox to\wd0{\hss\sl/\/\hss}}
\begin{document} 

\begin{center} {\Large\bf Covariant Symmetry Classifications for Observables of
Cosmological Birefringence}

\vskip 0.25in 
{John P. Ralston}

{\it Department of Physics and Astronomy, Kansas University, Lawrence, KS-66045,
USA\\ ralston@kuphsx.phsx.ukans.edu}

\vskip 0.25in 
 {\it and}
\vskip 0.25in 

{Pankaj Jain}

{\it Physics Department, I.I.T. Kanpur, India 208016\\ pkjain@iitk.ac.in}
\end{center}

\begin{abstract}
Polarizations of electromagnetic waves from distant galaxies are known to be
correlated with the source orientations. These quantities have been used to
search for signals of cosmological birefringence. We review and classify
transformation properties of the polarization and source orientation
observables.  The classifications give a firm foundation to certain practices
which have sprung up informally in the literature.  Transformations under
parity
play a central role, showing that parity violation in emission or in the
subsequent propagation is an observable phenomenon.  We also discuss
statistical
measures, correlations and distributions which transform properly and which can
be used for systematic data analysis.
\end{abstract}

\section{Introduction}
Galaxies at cosmological distances are known to emit polarized electromagnetic
radiation according to certain statistical regularities. 
The processes range from linear polarization from
radio galaxies due to synchrotron radiation, to centro-symmetric
polarizations sometimes seen in optical measurements of QSO's.
The
observables can include the percentage of polarization, a linear polarization
angle, a galaxy symmetry axis, and the redshift of the galaxy, as well as the
angular position coordinates of the galaxy on the dome of the sky.  One typical
correlation is an observed alignment of the bisector of a radio galaxy's long
axis and linearly polarized radiation in the ``FR1" class of objects.  This has
been interpreted as probing the source's magnetic field in conventional models.
While this effect is in the context of conventional physics, other polarization
correlations can test extremely small effects that might indicate new
physics--sensitive to ``cosmological birefringence" -- that might occur in the
subsequent propagation. This has led to several interesting studies, from
seminal early work on 
propagation in parity-violating electro-weak media by Leo
Stodolsky  \cite{Leo}, to the discussion of ``optical activity of the
universe"
by Gabriel Karl and collaborators \cite{Karl}, to recent 
studies of possible cosmological anisotropy in propagation \cite{NR,JR}.

\medskip

Galaxy and polarization axes are labeled by angles, typically measured in
``degrees East of North" in a spherical -polar system of angular coordinates.
The galaxy major axis projected into the tangent plane is measured with an
angle
$\psi$. A linearly polarized electric field orientation in the same coordinates
is assigned polarization angle $\chi$ . The angles, however, do not represent
{\it vectors} in the rigorous sense, because a rotation of $\pi$ radians
returns
an axis to its original orientation.  These observables are described by
elements of the real projective group, which is not very familiar in physics.
In this paper we will investigate and then automate proper transformation
properties of such observables, and list quantities which are invariant under
different symmetries.  This provides a firm foundation for certain traditions
and informal uses of variables which have sprung up in the literature.  Rather
interestingly, the split-up of the three dimensional orthogonal group into
disconnected subgroups of different parity also occurs for observables
connected
with polarization.  One then has quantities which are even or odd under parity,
which extends the list of symmetries one can test using polarization data.

\medskip The physical motivation for this is as follows. Consider a linear
polarization correlated with an emission axis.  
From ``symmetry", statistics would naively be expected to show equal offsets
of polarization axis relative to galaxy axis for either a positive or a
negative sense of rotation.
The particular ``symmetry"
assumed here is {\it parity} symmetry.  While the emergence of parity as an
observable may be surprising, it should be intuitively clear that
observation of
the converse would allow one to extract a definition of right--versus --left
handed senses of rotation from a data set, breaking parity symmetry.  An even
more obvious violation of parity symmetry would be preference for one kind of
circular polarization over another. 
While one may not expect an object the size of a galaxy to break
parity symmetry in its emission characteristics, the possibility
constitutes a valid observable, subject to classification and study.
Another more interesting possibility, moreover, is to assume that objects do
respect parity symmetry in a statistical sense upon emission, while the
intervening medium can be tested for parity violation to an extremely fine
degree.  
Many models predict just
such effects \cite{sikivie,ni}.
We will
proceed by setting up the transformation properties, listing covariant and
invariant quantities, and illustrating the remarks by some statistical
quantities which embody the ideas for data analysis.

\section{\bf Setup}
Let us discuss transformation properties. We are not concerned with Lorentz
boosts, nor the usual Lorentz covariant characterization of radiation in terms
of the energy momentum tensor, etc.  Our focus is the ``little group" of angular
variables in the rest frame of the observer, and the peculiar fact that the
angles ${\chi}$ and ${\psi}$ actually observed give the orientation of
``sticks".  A rotation of just $\pi$ radians about the axis of propagation
makes
no observable effect. Due to this unusual transformation property, both the
electric field and the galaxy axis are conveniently represented by certain
tensors.

\medskip

Let $\hat p$ be the radial unit vector pointing toward a particular source.
Construct a local Cartesian frame with its $z$--axis oriented along $\hat p$
(Fig 1).  The 2--dimensional subspace orthogonal to $\hat p$ is a plane tangent
to the unit sphere of the sky. Any consistent orientation of the local
$\hat x $
and $\hat y$ axes spanning the tangent plane can be used; one can always take
these to be in the directions of increasing azimuthal and polar angles ($\hat
\phi$ and $\hat \theta$) relative to a global North pole (hence the
astronomer's
``East of North").  The electric field $(E_x, E_y, 0)$  is transverse to $\hat
p$ and conventionally represented by a 2 component complex doublet, a
convention
hearkening back to optics \cite{mandle}.  In the circular polarization basis,
$|E>={1\/\sqrt{2}}(E_x+iE_y,E_x-iE_y)$.  The doublet notation is a hybrid
and the
transformation properties need examination.

\medskip

First, in averaging over many cycles of the wave, the observer does not
actually
measure the electric field. Instead what can be measured is an Hermitian
density
matrix \cite{mandle} $E_i E_j^* =J_{ij} =(J_{unpolz}+J_{polz})_{ij}$.  In our
local coordinates, the non--zero entries of $J$ are a $2 \times 2$ block in the
upper left hand corner of a $3 \times 3$ matrix.  If the coordinate system is
rotated, this tensor will transform in the usual way, because $E_i$ and $E_j$
transform like 3-vectors.

\medskip

The unpolarized part $J_{unpolz}$ is proportional to ``${\bf 1}$" with a
proportionality constant fixed by the total power.  The polarization density
matrix $J_{polz}$ is defined \cite{mandle} to be the rest of the matrix, ``as
if" it were a pure state.  Thus $det(J_{polz})=0$. This matrix is
conventionally
expanded in Pauli matrices:

$$ J_{polz}={1\over 2} P_{max} (1 +\hat\chi\cdot\vec \sigma) ={1\over
2}P_{max}(1 +\slash\chi\ ) $$ 
where $P_{max}$ is the degree of polarization,
$\hat\chi$ is a unit--3 vector, and $\vec\sigma$ are the Pauli matrices.
Equivalent are the ``Poincar\`e" coordinates $\vartheta_p$ and $\chi$ (for
describing elliptical and circular polarization \cite{mandle}):
$$\slash\chi=\pmatrix{\cos\vartheta_p&\sin\vartheta_p\; e^{2i\chi}\cr
\sin\vartheta_p\; e^{-2i\chi}& -\cos\vartheta_p\cr} $$
We have introduced the common ``slash" notation for contraction of a vector
with
the Pauli matrix. We can also use $\slash\chi$ for the $3 \times 3$ matrix when
no confusion between the two is possible. The polarization parameters have now
been organized in a form convenient for transformations.

\medskip

Consider the important case of a rotation of tangent plane coordinates
about the
propagation axis, that is the local z-axis, $|E>\rightarrow U| E>$, with
$U$ $3\times 3$ and in $SO(3)$. $U$ takes the block form $U=(u_\phi, 1)$
where $u_\phi$ is $2\times 2$ and unitary, in fact an element of SO(2). Of course
$\slash \chi \rightarrow u_\phi\slash\chi u^\dagger_\phi$.  This is the same
rule as the rotations in the spin 1/2 representation, but in this case
operating
on vector components, not spinors. It follows that the parameters $\hat \chi$
transform with the angle of rotation {\it doubled}.  As a check, a short
calculation shows that a linear polarization oriented at angle $\chi$ relative
to the local $x$--axis gives $\hat\chi=(\cos 2\chi,\; \sin 2\chi, 0)$ in the
local basis.  The factor of ``2" is just right to account for the
periodicity of
the observable plane of the electric field under rotations by ${\pi}$ (as
opposed to ${2\pi}$).  This supports the informal use of $\hat\chi$ in
literature ranging from biology \cite{batschelet,fisher} to astronomy
\cite{radio astron}, where the ``2" is inserted intuitively and to make
things come out right while making it clear that $\hat\chi$ is not a true
vector.

\medskip

The result is elucidated by the following general argument.  Recall the
familiar
decomposition of angular momentum ($\vec j$) states for the product  $ | E >
\times <E |  ={\bf 1 \times  1}= {\bf 2 + 1+ 0}$.  The states are labeled by
$(m, j)$ representing the eigenvalues of $j_z$ and $\vec j^2$, respectively.
However, since $E$ is transverse, all tensor products made from $m=0$ are
absent.  From the addition rule for $m$, these absent representations are the
$m= 0, m=\pm 1$ states of $j=2$, and the $m=\pm 1$ states of $j=1$. This
eliminates 5 of 9 possible combinations, leaving 4 total which are $ (2, 2),
(-2,
2), (0,1), (0, 0)$.  The 4 possibilities are the 4 Stokes parameters, with the
total power $\vec E \cdot  \vec E^*$ being the singlet $(0,0)$.  
The object $\hat\chi$, a deceptively vector--like position on the
Poincar\`e sphere, 
is made from the peculiar
combination $(2, 2), (-2, 2), (0,1)$.  Making a short calculation in the
helicity basis, $i P_{max} \cos(\vartheta_p) =1/2  Tr [ \sigma_z J] =
i(E_+E^*_+
- E_-E^*_-) = (\vec E \times\vec E^*)_z$ is the $(0,1)$ component, showing that
the ellipticity transforms like a 3-vector attached to the photon's
direction of
motion.

\medskip

A similar treatment is needed for a galaxy axis. An observed galaxy axis is the
projection of the physical major axis (a signless eigenvector of a $3 \times 3$
tensor of intensity distribution, say) onto the tangent plane. We can again
define in a helicity basis a 2 component $|\psi>=(\cos\psi+i\sin\psi,
\cos\psi-i\sin\psi)/\sqrt{2}$.   Again $|\psi>$ is not a good variable, because
the observable is actually projective, and we must identify $|\psi>=-|\psi>$.
The associated $2\times 2$ matrix, $|\psi><\psi|={1\over 2}+{1\over 2}\slash
{\bf \psi}$, is part of a tensor embedded in a $3\times 3$ matrix.  In our
coordinates it can be expanded in Pauli matrices
$$ |\psi><\psi|=1/2\pmatrix{1&e^{2i\psi} 
\cr h.c.& 1 \cr} ={1\over
2}{\bf 1}+{1\over 2}\slash{\bf \psi} $$
The matrix $\slash{\bf \psi}$ transforms like $u_\phi{\bf\slash \psi} 
u_\phi^\dagger$ and
defines a 3 component thing, $\hat\psi=(\cos 2\psi, \sin 2\psi, 0)$. 
Note again
the doubling of angles in $\hat\psi$, representing the information that
$|\psi>$ and $-|\psi>$ are now identified, just as informally practiced in the
literature.  (It is interesting that if one could deduce information on the
``pitch" of a galaxy, then $\vec\psi$ could point out of the plane: such an
axis
might be ``elliptically polarized", otherwise $\hat\psi$ is in the local
tangent
plane.)

\medskip

We also have the position on the sky $\hat p$, which is a true vector and {\it
not} to transform with doubled angles.

\medskip

Finally we should classify the objects under parity: by definition we have that
$\hat p \rightarrow -\hat p$.  $\hat\psi$ and $\hat\chi$ are only a little more
work: since $\hat\psi_j = 1/2 Tr[ \slash \psi \sigma_j]$, with $\slash \psi$
invariant, then $\hat\psi$ and $\hat\chi$ are invariant under parity (although
transforming in a complicated way under the full $O(3)$).

\medskip

Now one can transform away from the special coordinate system, using general $3
\times 3$ elements of $SO(3)$, with the three objects $\slash \chi$, $\slash
\psi$, and $\hat p$ all transforming properly.

\section{Covariants and Invariants}
We now turn to invariants one can make from the matrices $\slash \psi, \slash
\chi$ and $\hat p$.  A ``local" quantity will be one made from a single source,
or (if possible) different sources at the same location on the dome of the sky;
a ``non-local" quantity anything made from sources at different locations.

\medskip

{\it Local quantities:} Since for any galaxy $i$ a coordinate system exists
where all the matrix elements are in the upper left, then we have the covariant
identity for such $3\times 3$ matrices \begin{equation} \slash A\slash B =
A\cdot B\ 1 + i {\slash C } ; C= A\times B \end{equation}
The ``1" of course means the 2 x 2 unit matrix on the upper left, covariantly
written as $\delta_{ij} -\hat p_i\hat p_j$.  As a consequence of this identity
an obvious invariant is reduced to a simpler form: $$ s_1 = {1\over
2}Tr[\slash\chi\slash{\psi}]={1\over 2}Tr[\slash{\psi} \slash \chi] =
\hat\chi\cdot\hat\psi.$$ It follows that $s_1 = \cos(2(\chi-\psi))$, which is
clearly invariant under the local rotation $\chi \rightarrow \chi + \delta,
\psi
\rightarrow \psi + \delta$.  It also follows that $s_1$ is even under parity.

\medskip

Another useful quantity is the anti-symmetric $3 \times 3$ commutator $[\slash
\chi, \slash \psi ]$.  This is dual to a pseudo-vector:
\begin{equation} \vec A_i = 1/2 Tr [ \epsilon_i \slash \chi \slash{\bf \psi}]
\end{equation}
where $\epsilon_i$ is the completely anti-symmetric matrix with $j, k$ elements
$\epsilon_i^{jk}$.  In our local coordinate frame, $\vec A_i$ points in the
direction of $\hat p$ and is proportional to $\sin( 2 (\chi-\psi))$.  The sign
of proportionality depends on the right--handed convention for angles.  This
remains true however the coordinate system is rotated. The epsilon-tensor is
even under parity, making it clear that $\vec A$ is even and therefore a
pseudo-vector by construction. Since any quantity which is odd in $(\chi-\psi)$
reverses when a positive ``sense" of rotation is reversed to a negative one,
such quantities are {\it parity--odd} on general grounds.

\medskip

This is not the only odd-parity observable.  The helicity $h$ of the wave
is defined in a Lorentz-covariant manner as the projection of its spin
along its direction of propagation.  This is a pseudo-scalar: it must
therefore be equal (up to a constant) to the product $\hat\chi\cdot\hat p$.
The other invariants which can be made by contracting $\hat p$ and the matrices
are trivially zero.  Consulting the identity (1) above, there are 4 real-valued
quantities in the products of the 2 matrices, which have now been classified
into one scalar and one pseudo-vector, exhausting the possibilities for local
bilinears of the two tensors. One can go further, and add another unit vector
$\hat\lambda$ to the problem. Such a vector is needed to quantify
asymmetries of
angular distribution. Then one can make a scalar $s_2$ and a pseudoscalar
$p_1$:

$$ s_2= \hat p\cdot\hat\lambda = \cos( \theta);\; p_1= \vec
A\cdot\hat\lambda=\sin( 2 (\chi-\psi)) \cos( \theta) $$ where $\theta$ is the
angle between $\hat\lambda$ and the position of the source on the dome of the
sky. The pseudovector combination $\hat p \times \hat\lambda$ can also be
considered, which makes the usual unit vector $\hat \theta$ sitting in the
tangent plane, transforming like $(\pm 1, 1)$.  
By the $j_z$ addition rule for angular momentum, this
cannot be combined with the $(2,2), (-2,2), (0,1) $ representations
available from $\slash \chi$ and $\slash \psi$ to make new
invariants.  Using $\hat p$, which transforms like $(0,1)$, we can take the
$(0,1)$ part
of $\slash\chi$ and make a pseudo-scalar, but this is the helicity $h$
already discussed.

\medskip

Continuing in this way, products of higher order can always be reduced to sums
of smaller dimensional representations, much like the usual decomposition of
angular momentum.  Some care is needed, however, because the polarization
variables are made from incomplete representations because the fields are
transverse.  One generally then has fewer invariants than straightforward
counting using rotation group methods might indicate.

\medskip

{\it Non-Local Invariants:} Combining different sources leads to some
interesting quantities at low order.  There is for example the familiar angle
$\theta_{ij}$ between galaxy $i$ and galaxy $j$ on the sky, given in terms of
the true scalar $cos(\theta_{ij}) =\hat p_i\cdot\hat p_j$.

\medskip

More interesting are symmetric and antisymmetric combinations of $\slash\chi_i$
and $\slash\chi_j$, or $\slash\psi_j$. It is straightforward to treat these by
standard methods: thus the symmetric combinations $$ Tr[ \slash\chi_i
\slash\chi_j] ;\; \;  Tr[ \slash\chi_i \slash{\psi}_j] ;\; \; Tr[
\slash{\psi}_i
\slash\chi_j] ;$$ are scalars; while the antisymmetric combinations or
commutators $$ [\slash\chi_i , \slash\chi_j] ;\; \; [\slash\chi_i ,
\slash\psi_j] ;\; \; [\slash\psi_i , \slash{\psi_j}] ;$$ are dual to
pseudo-vectors.

\medskip

The quantities above are involved in questions of angular coherence: for
example, if one wants to smooth the angular distribution of sources at
different
locations, while it would be a bit sloppy to add the parameters such as
$\hat\chi$, in practice we are aware only of averaging polarization parameters
over single sources of small angular dimensions, in a quasi-local way, so that
this should not cause a problem. 
However if one were making a more ambitious study --for example examining
the degree of coherent polarization over larger regions of the sky--then
more formal care would be needed.  One can, for example, create correlation
functions of the matrices in a consistent basis and evaluated at different
angular positions, which are then reduced to scalars by taking traces.
Another interesting quantity is the $h_i h_j$, or helicity-helicity
correlation function, and obvious generalizations removing the means.  This
quantity does not need a galaxy axis for its evaluation, and might probe
parity-violating effects of the medium evaluated as a function of angular
scales.  This may be useful for optical polarizations , for example, which
often are not associated with axial structure.  An application of interest
is the cosmic micro-wave background, for which polarization measurements
can be expected in the future.

\medskip
\noindent
{\it\ Distributions} 

The classifications are useful in constructing statistical
distributions or correlations which can be used to quantify observations.  We
will use a bracket ( $<  >$ ) symbol to denote an expectation value in a
normalized distribution.

\medskip

Consider the problem of quantifying the correlations between $\chi$ and $\psi$
mentioned in the Introduction. From our results, one would naturally assume the
distribution to be a function of the rotationally invariant quantity
$\beta=\chi-\psi$.  Given that $\chi$ and $\psi$ are defined up to multiples of
$\pi$, the difference $\beta$ ranges in the most general case over $2 \pi$, and
not $\pi$, as sometimes assumed \cite{CF} in the literature.  This is seen
very simply by making sketches of some trial distributions. Analytically
one can
expand the distribution $f( \chi, \psi)$ in Fourier series for integer $n, m$:
$$ f( \chi, \psi) = \sum  f_{nm} \exp ( i 2n \chi+ i 2 m \psi) =\sum f_{nm}
\exp
\left[ i (n - m) ( \chi- \psi)  + i (n+ m) ( \chi+\psi)\right] $$ 
Reality of $f$ must also be imposed; it prescribes the negative
integer values of $f_{nm}$, but does not restrict whether
$n \pm m$ is even or odd. The coefficients $f_{nm}$ for odd values of $n - m$
make $\beta=\chi-\psi$ periodic on the interval of $2\pi$, as claimed.
Restricting non-zero coefficients to even values of $n - m$ can be
motivated by extra assumptions.  One sufficient condition is that the
distribution obeys overall rotational symmetry, giving $f_{nm}$ going
like $\delta( n+m) $.  Such an assumption might seem very general but in
fact it is not.  
It is an interesting fact of optics that even a perfectly
transmitting (unitary) medium can treat different polarizations
dissimilarly, leading to a non-trivial distribution of $\chi+\psi$ as a
signal. 
The same mechanism can
create an anisotropic distribution of linear polarizations along a pencil
through a medium from an random uncorrelated  distributed set of linear
polarization emitters. 
To simplify the analysis we will assume here, however, that the
distribution of $\beta$ is periodic on the interval of $\pi$.

\medskip

To make invariant distributions from invariants, this leaves us with the
quantities constructed earlier, namely $s_1=\cos(2\beta)$ and $p_1=\sin(2
\beta)$.  A simple, and indeed well-known distribution that follows is the von
Mises form \cite{batschelet,fisher,mardia}: $f_{vM}(\beta) = const.\
\exp(
k \cos(2\beta)) = const. \exp( k s_1 ) $.  This distribution has often been
used
in likelihood tests, but unfortunately without discussion of parity symmetry.
By construction, the $vM$ distribution embodies a physical assumption that the
twist of one angle relative to another has no preferred parity, which may be
unsuitable in some cases.  
The ``shifted" von Mises distribution is similar: $f_{\rm shifted-vM}
(\beta) = {\rm const.} \exp(k \cos(2(\beta- \alpha))$.  
The parameter $\alpha$
shifts the origin of $\beta$ and might seem to be free.  However if parity
symmetry is assumed, then $\alpha$ is quite restricted.  The distribution is
a function of $\cos(2\beta- 2\alpha)=\cos(2\beta) \cos(2 \alpha) +
\sin(2\beta)\sin(2 \alpha)$.  Parity symmetry requires 
$f(\beta)=f(-\beta)$,
yielding $\sin(2\alpha)=0$, or $\alpha=0, \pi/2$.  
Parity symmetry, then, is
sufficient to predict that marginal distributions of linear polarizations
tend to be oriented either along galaxy major axes, or perpendicular.  This
has been a misunderstood point, because of assertions that only the single
choice of angular origin relative to the perpendicular is sensible.  For
any statistic which is covariant under a shift of origin, however, the
matter is irrelevant, making no difference.  Transforming
$\beta\rightarrow\beta - \pi/2$ is equivalent to transforming
$k\rightarrow -k$ in the $vM$ or shifted $vM$ distributions, showing that
either choice is equally well described automatically.  
Similarly, $\sin(2
\beta)$ is odd in both $\beta$ and $\beta- \pi/2$, a pseudoscalar no
matter how angles are measured.  Another example of the issue is
underscored by the paper of Loredo et al.\cite{loredo} using the shifted
$vM$ distribution in a detailed likelihood analysis.  The paper responded
to an odd-parity statistic used in Ref. \cite{NR}, but exclusively used
an even parity shifted $vM$ model in its analysis.  In replacing the
original statistic by one with opposite transformation properties,
conclusions were drawn on a false basis.  This illustrates the problem that
can occur when using functions from a different symmetry class than the
idea being tested.

\medskip

There is in fact a long history of mix-ups from lumping together observables of
different parity, which has led to interesting contention in the literature.
Birch \cite{birch} in 1982 empirically observed a {\it pseudovector}
correlation in radio polarizations, 
odd in the difference $(\chi-\psi)$.  Kendall and Young ($KY$)
made
a model distribution to explore this\cite{KY}, choosing $vM$ for the
{\it null} distribution of $\beta$. This is an implicit assumption of parity
conservation in the null, which is perfectly physical (but which should be
stated explicitly).  $KY$ then set up a conditional correlation function
$C_{KY}( \beta, \hat p)=exp( \mu \hat\lambda\cdot\hat p \sin(2\beta))$ which
involved odd-parity to test Birch's pseudovector result.  
Unfortunately Kendall and Young did not explain the reason
for their ansatz, which in retrospect was entirely appropriate.
The overall $KY$ distribution finally can be written
in a
nicely compact form $$ f_{KY}(\chi, \psi, \hat p) =exp( k\cos( 2\beta)
+\mu\hat \lambda\cdot\hat p \sin(2\beta)) =\exp( k s_1 + \mu p_1) $$ where $s_1$
and $p_1$  are our scalar and pseudoscalars from the first section, making it
clear that $\mu$ is a parity-violation anisotropic correlation parameter.
Likelihood analysis, for example, can be used in a perfectly objective way to
see whether any parity--violating effects might be present or not in a data
set,
and entirely separate from the need to quantify the marginal distribution of
$\beta$.  It is interesting that Kendall and Young's data analysis \cite{KY}
along such lines then indicated a high statistical significance for the
parity
-violating effect Birch had observed; more recent work in this regard can also
be cited \cite{JR}. However, other independent studies which used invariant
correlations lumping together different parity seemed to contradict the result,
a topic to which we now turn.

\medskip
\noindent
{\it\ Correlations} 

One of the best-known statistical correlation tests comes
from an influential paper by Jupp and Mardia\cite{JM}. Their prescription
for correlation between 2 angular quantities involves mapping them into
''vectors" $v, w$, that is covariantly transforming elements, of which
$\hat\chi$
and $\hat\psi$ are examples we have already seen.  To test for correlations,
$JM$ use canonical $p \times q$ correlation matrices $$\Sigma^{vw}_{ij} = <
(v_i-<v_i>)( w_j-<w_j> )> $$ which behave under the separate transformations of
$v, w$ as the indices would indicate.  Then the $JM$ correlation test is to
calculate $$ n\rho^2_{v, w} = n\ Tr[\Sigma^{vw} (\Sigma^{ww})^{-1}
\Sigma^{wv}(\Sigma^{vv})^{-1} ], $$
with matrix products indicated, and where $n$ is the number of data points in
the sample.  The step of dividing by auto-correlation matrices is used to make
the quantity scale-invariant.  One easily finds that $n\rho^2_{vw}=0 $ if $v$
and $w$ are independent, and the distribution of fluctuations in an
uncorrelated
null distribution has also be obtained\cite{JM}.  
Thus $n\rho^2_{v, w} $ has served for many
years as a useful simple test for independence.

\medskip

Unfortunately Jupp and Mardia, in discussing their correlation test, did not
discuss parity and other discrete symmetries. In the problem at hand, to test
whether the difference $\beta$ is correlated with sky position, one might
consider the $JM$ correlation of a ``natural vector" $v=(\cos(2\beta),\sin
(2\beta))$ and $w= \hat p$.  This particular combination $v$ was used to test
Birch's correlation by Bietenholz and Kronberg\cite{BK}. 
Recall, however, the even-odd rule of
parity for those quantities even-or odd in $\beta$. It is clear that
the above "natural" vectorlike combination $v$ needlessly mixes two
quantities which are of opposite parity.  The two pieces are also
separately invariant under rotations, and should not be combined
into a vector.  
  Ironically, history shows that the mixed parity combination was used
while at the same time citing that Birch had a correlation of pseudovector
character \cite{BK}; without recognizing the parity properties of Kendall and Young's
procedure, both their and Birch's results were then rejected. From our analysis
it is sufficient to use $n\rho^2_{v_1, w} $ and  $n\rho^2_{v_2, w} $, which
allows separate tests of scalar or pseudoscalar kind. One can, of course,
also work directly with scalar and pseudoscalar measures such as $<s_1>$ and
$<p_1> $ to explore certain features of data, and create any number of
statistics, once the proper transformation properties have been respected.

\section{Summary}
We have classified several combinations of the polarization and angular
correlation observables under parity and angular momentum.  The features of
local observables, and their simple appearance in classic distributions and
correlation analysis, should be helpful to those interested in the area. While
not exploring very far into the non-local quantities, they appear to offer many
possibilities for interesting studies. The possibility of exploring parity
violation when there are no obvious emission axis variables, as in the case of
upcoming polarized cosmic -microwave background observations from ground-based
and satellite facilities, seems intriguing and bears further investigation.

\medskip

ACKNOWLEDGMENTS: This work was supported in part under DOE grant number
85ER40214, by the University of Kansas General Research Fund, and the {\it
Kansas Institute for Theoretical and Computational Science/ K*STAR} program.

\begin{figure}
\psfig{file=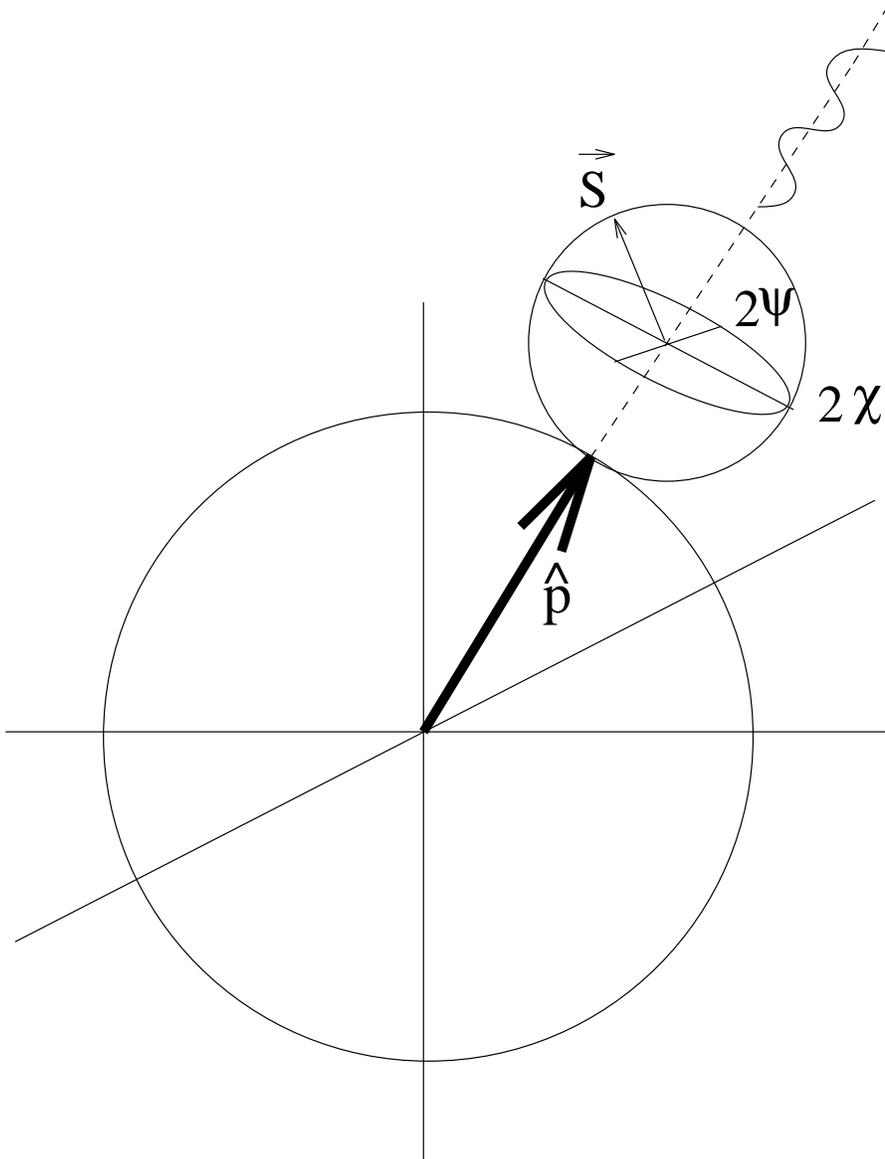}
\caption{
 Schematic diagram of observables.   The variables $\chi$ and
$\psi$ are orientation angles of  the plane of electric field linear
polarization, and of a source symmetry axis, respectively.   A unit vector
pointing towards the source on the dome of the sky (big sphere) is denoted
by  $\hat p$.  The polarization density matrix assigns a standard quantity
$\vec s$ on the Poincar\`e sphere (small sphere).  The projection of $\vec
s$  in the tangent plane transforms like spin-2 under rotations about the
local $\hat p$ axis;  the component of $\vec s$ along $\hat p$ is a measure
of ellipticity, and transforms under a spin-1 representation.
}
\end{figure}

 \end{document}